\documentclass[aps, prl, twocolumn, superscriptaddress, showpacs, floatfix]{revtex4-1}
\usepackage{amsmath,amsbsy,amssymb,amsfonts, stackengine}
\usepackage{microtype}

\usepackage{graphicx, color}
\usepackage{natbib}
\usepackage{hyperref}
\usepackage[capitalise]{cleveref}
\usepackage{subfigure}
\usepackage{float}

\usepackage{soul,xcolor}
\usepackage{marginnote}
\newcommand{\Matr}[1]{\boldsymbol{\mathcal{\hat{#1}}}}
\newcommand{\eps}{\epsilon}
\newcommand{\vc}[1]{\pmb{#1}}

\graphicspath{{Figures/}}
\begin{document}
\date{\today}

\title{Optical Kinetic Theory of Nonlinear Multi-mode Photonic Networks}

\author{Arkady Kurnosov}
\affiliation{Wave Transport in Complex Systems Lab, Department of Physics, Wesleyan University, Middletown, CT-06459, USA}
\author{Lucas J. Fern\'andez-Alc\'azar}
\affiliation{Institute for Modeling and Innovative Technology, IMIT (CONICET - UNNE), Corrientes W3404AAS, Argentina}
\affiliation{Physics Department, Natural and Exact Science Faculty, Northeastern University of Argentina, Corrientes W3404AAS, Argentina}
\author{ Alba Ramos}
\affiliation{Institute for Modeling and Innovative Technology, IMIT (CONICET - UNNE), Corrientes W3404AAS, Argentina}
\affiliation{Physics Department, Natural and Exact Science Faculty, Northeastern University of Argentina, Corrientes W3404AAS, Argentina}
\author{Boris Shapiro}
\affiliation{Technion -- Israel Institute of Technology, Technion City, Haifa, 3200, Israel}
\author{Tsampikos Kottos}
\affiliation{Wave Transport in Complex Systems Lab, Department of Physics, Wesleyan University, Middletown, CT-06459, USA}
\date{ }

\begin{abstract}
Recent experimental developments in multimode nonlinear photonic circuits (MMNPC), have motivated the development 
of an optical thermodynamic theory that describes the equilibrium properties of an initial beam excitation. However, a non-
equilibrium transport theory for these systems, when they are in contact with thermal reservoirs, is still {\it terra incognita}. 
Here, by combining Landauer and kinematics formalisms we develop an one-parameter scaling theory that describes the 
transport in one-dimensional MMNPCs from a ballistic to a diffusive regime. We also derive a photonic version of the 
Wiedemann -Franz law that connects the thermal and power conductivities. Our work paves the way towards a fundamental 
understanding of the transport  properties of MMNPC and may be useful for the design of all-optical cooling protocols.
\end{abstract}

\maketitle
{\it Introduction --}Quite recently, we have witnessed a surge in understanding and harnessing the convoluted behavior of light 
propagation in multimode nonlinear photonic circuits (MMNPC). These experimental platforms provide an excellent testbed not 
only to investigate but also unravel new horizons and possibilities with both fundamental and technological ramifications. For 
example, they have been used for exploring exotic optical phase transitions \cite{SF20,KSVW10,S12,RFKS20} and beam self-
cleaning phenomena \cite{KTSFBMWC17,LWCW16,N19}, spatiotemporal mode locking \cite{WCW17}, multimode solitons 
\cite{WCW15}, etc. In parallel, their implementation in fiber-optical communications might resolve urgent technological needs 
associated with the looming information “capacity crunch”  \cite{HK13,RFN13} or the quest for new platforms of high-power 
light sources \cite{WCW17}. 

Nonlinearities lead to multi-wave mixing processes or photon-photon “collisions” through which the many modes can exchange 
energy via a multitude of possible pathways, often numbering in the trillions even in the presence of one hundred  modes or so. 
Evidently, modeling, predicting, and harnessing the response of such exceedingly complex configurations is practically impossible 
using conventional brute-force time-consuming computations that obscure the underlying physical laws. Fortunately, an entirely 
new universal approach inspired by concepts from statistical thermodynamics emerged recently \cite{Wu2019}, that self-consistently 
describes the utterly complex processes of energy/power exchange in MMNPC. Under {\it thermal equilibrium} conditions, the 
methodology has  identified intrinsic variables (optical temperature $T$ and chemical  potential $\mu$) that play the role of optical 
thermodynamic forces leading an initial beam excitation to a Rayleigh Jeans (RJ)  thermal state \cite{Wu2019,RFKS20,MWJC20} 
-- a key tenet of this theory that has been confirmed using multimode optical fibers and time-synthetic photonic lattices  \cite{PSWBWCW22,MWJKCP23,BGFBMKMP23}.

Here, we develop a transport theory that describes the {\it nonequilibrium} steady states generated in a MMNPC (a photonic junction) 
when is in contact with two optical reservoirs at different optical temperature and/or chemical potentials. The proposed optical kinetics 
framework, allows us to define and calculate various kinetic coefficients like optical power and thermal conductivities in full analogy 
with physical kinetics in condensed matter \cite{LL10}. The presence of two conserved quantities (total optical power and electrodynamic 
momentum flow, referred below as internal energy) requires to consider the coupling between thermal and power currents mediated 
by nonlinear interactions -- a complication that is not present in phonon heat transport in solids. Using a combination of Landauer 
(ballistic limit) and Boltzmann (diffusive limit) transport theories, we established a one-parameter scaling theory that describes the 
crossover from a ballistic to a diffusive limit as the size of the photonic junction increases beyond a characteristic length-scale $l_T$ 
that incorporates information about the thermalization processes occurring at such scales in the junction. Finally, we analyze the 
interdependence of optical power and heat transport by deriving the photonic analogue of Wiedemann-Franz (WF) law that connects 
the thermal and power conductances. Their ratio is inversely proportional to temperature -- as opposed to the linear temperature 
enhancement occurring in typical thermoelectric devices -- which is a signature of relaxation scales separation between the energy and 
power currents. Our results set the basis for the development of novel thermo-photonic devices and paves the way for the design of novel 
photonic refrigerators or engines.

\begin{figure}
\center\includegraphics[width=\columnwidth]{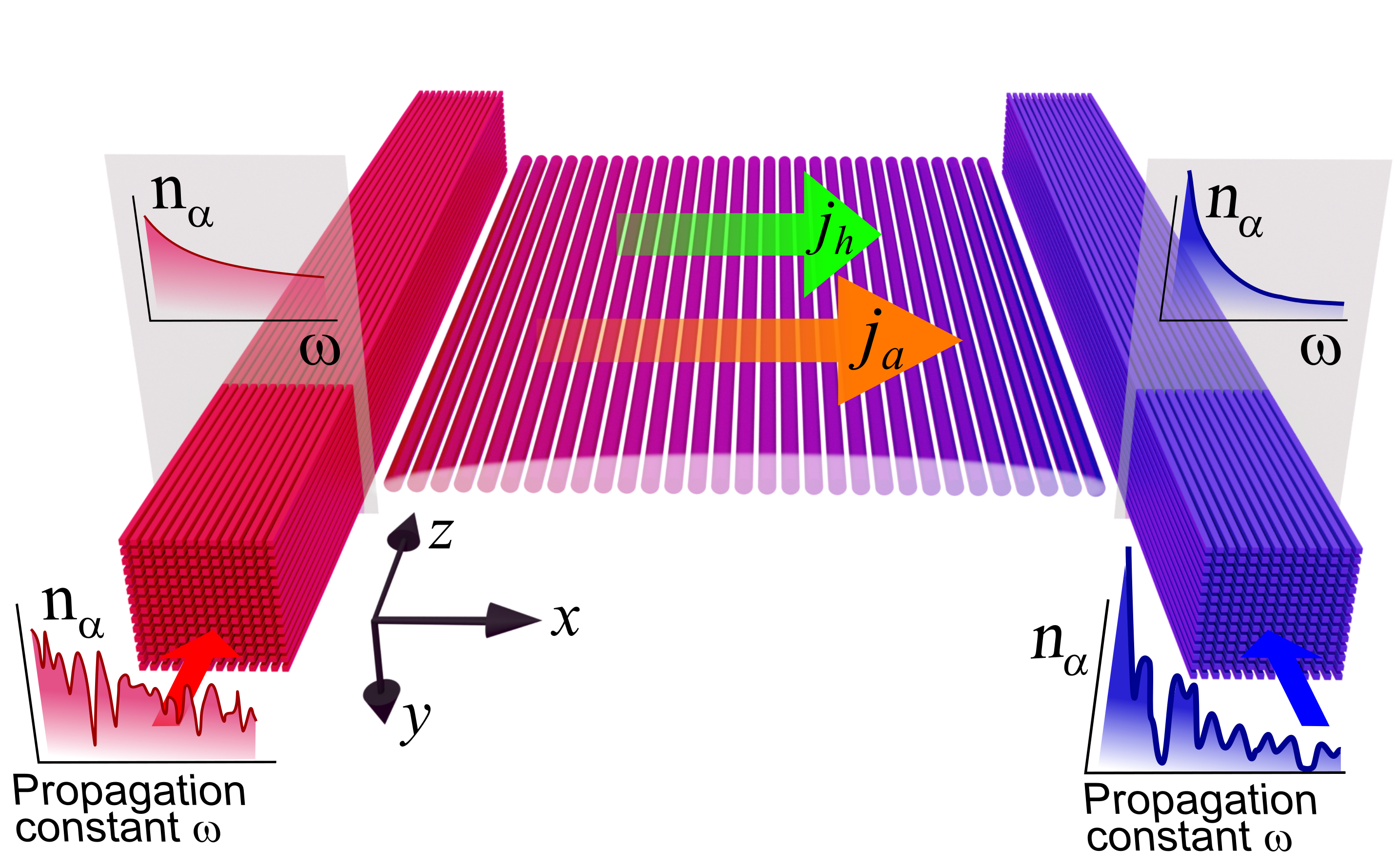}
\caption{ The two photonic reservoirs (L/R), consist of a large number of coupled nonlinear 
waveguides. We initially inject two beams at each of the reservoirs with different internal energy and power. At the steady 
state, the power distribution is given by a RJ with predetermined temperatures $T_{L/R}$ and chemical potentials $\mu_{L/R}$. 
After reaching a thermal state, the two reservoirs are coupled by an optical junction which transports heat and power currents. 
For intermediate, but large, length scales these currents acquire a quasi-steady-state $j_h, j_a$.
}
\label{fig1}
\end{figure}

{\it Physical setting -- }We consider the set-up shown in Fig. \ref{fig1}. It consists of two optical reservoirs and a junction which 
facilitates thermal and power transport between the reservoirs. The left (L) and right (R) reservoirs consist of arrays of (weakly) 
nonlinear coupled optical waveguides (or multimode/multicore optical fibers) supporting a finite (but large) number $\gamma=1,
\cdots, M$ of linear supermodes $\left|\phi_\gamma^{(L/R)}\rangle\right.$ -- all propagating along the axial direction $z$ with 
propagation constants $\omega_\gamma^{(L/R)}$. At each reservoir, we launch a beam prepared at some state $\left|\Psi^{(L/R)}
(z=0)\rangle\right.=\sum_\gamma C_\gamma^{(L/R)}\left|\phi_\gamma^{(L/R)}\rangle\right.$ where $C_{\gamma}^{(L/R)}=
\langle\phi_\gamma|\Psi^{(L/R)}\rangle$ are the projection coefficients to their supermodes. Initially, the reservoirs are decoupled 
from one another and, therefore, their total optical power $\mathcal{N}^{L/R}(\{C_\gamma^{(L/R)}\})=\sum_{\gamma} 
|C_\gamma^{(L/R)}|^2 \equiv A^{(L/R)}$ and internal energy $\mathcal{H}^{L/R}(\{C_\gamma^{(L/R)}\})\approx \sum_\gamma 
|C_\gamma^{(L/R)}|^2\omega_\gamma \equiv E^{L/R}$, are constants of motion which are used to determine the optical 
temperature $T $, and chemical potential $\mu$ that define their RJ thermal state $n_{L/R}(\omega_{\gamma}^{(L/R)}) = \langle
|C_\gamma^{(L/R)}|^2\rangle\equiv \frac{T_{L/R}}{\omega_{\gamma} -\mu_{L/R}}$ \cite{RFKS20,Parto2019,Shi2021,Wu2019}. 

Once each reservoir reaches a thermal equilibrium with $(T_L,\mu_L)\neq (T_R,\mu_R)$, they are coupled with an 
optical junction. It consists of an array of coupled (weakly) nonlinear single-mode waveguides supporting $\alpha=1,\cdots, N$ 
linear supermodes $\left|\phi_\alpha^{(J)}\rangle\right.$ that propagate along the paraxial distance $z$ with propagation 
constants $\omega_\alpha^{(J)}$.  

The junction facilitates heat and power transport between the reservoirs. Obviously, the characteristics of the junction, 
i.e., coupling between waveguides, nonlinearity strength, size, etc., will determine these currents, and, in turn, the paraxial 
length $z_{\rm GT.}$ at which the whole system reaches a global thermalization. We are not interested in the behavior 
of the system at these (practically irrelevant) large paraxial length-scales. Rather, we focus our investigation at a physically 
relevant intermediate (but still large) length scales $z_{\rm tr}< z\ll z_{\rm GT}$, where (after a transient paraxial length 
$z_{\rm tr}$) the currents through the junction acquire (quasi-)steady-state values. Our goal is to develop a non-equilibrium 
transport theory that describes such power and heat transfer at these intermediate length scales. 

{\it Mathematical modeling -- }The beam dynamics at the junction and the left/right reservoirs is described by a temporal 
coupled mode theory (TCMT), 
\begin{equation}\label{Eq:TCMT}
i\frac{d\Psi_{l}}{dz} =  - \sum_{j} J_{lj} \Psi_j + \chi|\Psi_{l}|^{2} \Psi_l,
\end{equation}
where $\Psi_l\equiv\langle l|\Psi\rangle$ is the field amplitude at the $l$-th waveguide, $\eps_{l} = -J_{ll}$ is its propagation 
constant and $\chi$ is the Kerr nonlinearity coefficient. The connectivity of the network is dictated by the coupling coefficients 
$J_{lj}^{} = J_{jl}^{\ast}$. At the junction $J_{lj} = J\delta_{l, l\pm 1}$ ($\epsilon_{l} = \epsilon = 0$). The corresponding 
linear dispersion relation takes the form $\omega_{\alpha}^{(J)}=-2J\cos(k)$ where the wavevector $k\in[-\pi,\pi]$. The two 
reservoirs consist of a square lattice of $N_r=40\times 40=1600$ waveguides with $J_{lj}=J$. To avoid spectral degeneracies, 
we consider propagation constants given by a uniform distribution $\epsilon_l\in [-\frac{W}{2},\frac{W}{2}]$ with $W=0.5$. 
The junction-reservoir coupling is $J_{b-r}=0.2 J$. We have confirmed via direct dynamical simulations of the composite system, 
the existence of a (quasi-)steady-state regime during which the temperature and chemical potential at each reservoir remain 
(approximately) constant when $M\gg N$. 

{\it Onsager Matrix Formalism -- }We consider the power $a(x)$ and energy $h(x)$ densities at any segment $(x, x+dx)$ inside 
the junction which includes many unit cells. At the (quasi-)stationary regime, these segments are at local equilibrium characterized 
by a local temperature and chemical potential, i.e.,  $T = T(x)$, $\mu = \mu(x)$ which slowly change with the position. Under this 
assumption, the currents are evaluated by expanding the spatial gradients of $\mu,T$ up to the first term \cite{Callen1985,
Saito2010}:
\begin{equation}\label{Eq:Onsager}
{\bf j}= \Matr{L} {\bf f}, 
\quad \Matr{L}=
\begin{pmatrix}
    {\cal L}_{aa} & {\cal L}_{a q}\\
    {\cal L}_{q a} & {\cal L}_{q q}
\end{pmatrix}.
\end{equation} 
Above, ${\bf j}=(j_a,j_q)^T$ where  $j_a, j_h, j_{q}= j_{h} - \mu j_{a}$ are the power, energy, and heat currents, $\Matr{L}$ 
is the the Onsager matrix, while the affinities ${\bf f}=(-\nabla\mu/T,\nabla (1/T) )^T$ are the thermodynamical forces that induce 
the currents. For systems preserving time reversal symmetry, the Onsager reciprocity relations hold, i.e., ${\cal L}_{q a}=
{\cal L}_{a q}$ \cite{Callen1985,Benenti2020}. 

We distinguish between two limiting cases of short $N\ll l_T$ (ballistic) and long $N\gg l_T$ (diffusive) junctions where $l_T =
v_g z_T$, $v_g$ is a typical group velocity of the linear supermodes, while $z_T=z_T(\chi,J,T,a)$ is a relaxation distance that 
dictates the thermalization process of a non-equilibrium state in the isolated junction towards its RJ distribution \cite{Shi2021,
Ramos2023}. Strictly speaking the concept of local equilibrium used in Eqs. (\ref{Eq:Onsager}) applies to the diffusive regime 
only while in the ballistic regime, the meaningful quantities are the temperatures $T_{L,R}$ and chemical potentials $\mu_{L,R}$ 
of the two reservoirs. In this case we can formally define $\nabla T\equiv (T_R-T_L)/N; \nabla \mu\equiv (\mu_R-\mu_L)/N$ to 
have unified notations for both regimes.

The various transport coefficients, can be extracted from the elements of the Onsager matrix \cref{Eq:Onsager}. For example, 
the power conductivity is $\sigma\equiv {\cal L}_{a,a}/T$, the thermal conductivity is $\varkappa = \det\Matr{L}/(T^{2}
{\cal L}_{aa})$, while the Seebeck and Peltier coefficient that describe thermo-power transport are $S = {\cal L}_{aq}/(T 
{\cal L}_{aa})$ and $\Pi=T\cdot S$ (see Supplement for details). Below, we analytically evaluate these matrix elements. 

\hspace{-0.4cm}\textit{(a) Ballistic regime --} In this regime, the nonlinear interactions are not able to enforce mixing among the 
linear modes. The power and energy fluxes through the junction are evaluated using Landauer's theory \cite{Imry1999} 
\begin{equation}\label{Eq:CurrentsLandauer}
j_{a (h)} = \int d\omega \,t(\omega)\omega^{s}\left[n_{\rm L}(\omega) - n_{\rm R}(\omega)\right],
\end{equation} 
where $s = 0$ ($s=1$) for power (energy) current. We further assume that the  transmittance $t(\omega)=t_0={\rm const.}$ for all 
supermodes in the band $[-2J,2J]$ and zero otherwise. Equation  (\ref{Eq:CurrentsLandauer}) can be evaluated analytically, thus 
allowing us to extract the Onsager matrix elements (see Supplementary Material)
\begin{equation}\label{Eq:OnsagerBallistic}
\begin{split}
&{\cal L}_{aa} = t_{0}N\frac{4J T^2}{\mu^2 - 4J^{2}}; \quad {\cal L}_{qq} = t_{0}N4JT^{2}\\
&{\cal L}_{aq} = {\cal L}_{qa} = -t_{0}NT^{2}\ln\left(1 - \frac{4J}{-\mu + 2J}\right),
\end{split}
\end{equation}
where $T = (T_{L} + T_{R})/2$, $\mu = (\mu_{L} + \mu_{R})/2$ and $|\Delta\mu|\ll |\mu| - 2J$ and $\Delta T\ll T$ (linear response 
regime). Equations (\ref{Eq:OnsagerBallistic}) are the main results of this section. They allow us to derive exact expressions for the 
transport coefficients $\sigma,\varkappa \propto N$. We, furthermore, conclude that when $l_T\gg N$, the Fourier's law does not hold.

While the weak nonlinear interactions cannot enforce sufficient mode-mode mixing, they can induce a nonlinear frequency shift 
$\omega_{\alpha}^{(J)}\to \omega_{\alpha}^{(J)} + 2 a\chi$ (see Supplement, and Ref. \cite{Basko2014}) which might affect 
the value of the currents. Nevertheless, \cref{Eq:CurrentsLandauer} still applies with the modification that the transmittance acquires 
its constant value $t_0$ inside a {\it shifted} frequency window $\omega^{(J)}\in [ -2J+2\chi a, 2J+2\chi a]$ while it is zero everywhere 
else. As a result, \cref{Eq:OnsagerBallistic} still applies with the substitution $\mu_{L, R}\to \mu_{L, R} - 2a\chi $. This nonlinear 
frequency correction $j_{a(h)}\rightarrow j_{a(h)}^{(\chi)}$ is insignificant in the high temperature limit, $T\sim|\mu|\gg 2a\chi$ 
but it becomes important when $|\mu|\sim 2J$. 

\hspace{-0.4cm}\textit{(b) Diffusive transport --} In the other limiting case of $N\gg l_T$, the nonlinear mode-mode interactions become 
a dominating mechanism of power and heat transport. They are responsible for a local equilibrium within 
$l_T$ segments of the junction, thus allowing us to define slowly varying local temperatures $T(x)$ and chemical potentials $\mu(x)$. 
At the same time, the modal occupations also become a local quantity, i.e., a function of coordinate $x$, wave vector $k$, and propagation 
distance $z$, $n = n(x, k, z)$. We proceed by invoking a Kinetic Equation (KE) approach \cite{Basko2014} 
\begin{equation}\label{Eq:KineticEquation}
\frac{dn}{dz} = \frac{\partial n}{\partial z} + (v_g\cdot\nabla n) = \mathrm{St}\,n,
\end{equation}
where $n(k, x)dkdx$ represents the power (number of particles) contained in a macroscopic volume element $dkdx$ of the phase space 
and $\mathrm{St}\,n$ is a collision integral. Next, we consider the stationary regime $\partial n/\partial z=0$, and assume that $n$ 
depends on the position $x$ in the junction via the local optical temperature $T(x)$ and chemical potential $\mu(x)$. Further, we 
linearize \cref{Eq:KineticEquation}, assuming small deviations from the local equilibrium, $n = n^{(0)} + \delta n$, where $n^{(0)}$ 
is the (local) equilibrium RJ-distribution. Since $\mathrm{St}n^{(0)}=0$, the rhs of \cref{Eq:KineticEquation} becomes $\mathrm{St}
\,\delta n \approx -\delta n/z_T(k)$ (``time''-relaxation approximation). Then, the solution of the linearized KE reads
\begin{equation}\label{Eq:KineticLinear}
\delta n(k) \approx -z_T(k)\frac{n^{(0)}}{T}\left[(v_g\cdot\nabla\mu)n^{(0)} + (v_g\cdot\nabla T)\right].
\end{equation}
resulting to power and heat currents
\begin{equation}
j_{a(q)} = \int\frac{dk}{(2\pi)}v_g(k)\left[\omega^{(J)}(k) - \mu\right]^s\delta n(k). 
\end{equation}
Evaluation of the above integrals [See Supplement for details] together with \cref{Eq:Onsager} allows us to express the Onsager 
matrix elements as   
\begin{eqnarray}
{\cal L}_{aa} &=& -T^{2}\left[1 + \frac{\mu}{\sqrt{\mu^{2} - 4J^{2}}}\right]z_T, \quad {\cal L}_{qq} = 2 J^{2}T^{2} z_T\nonumber \\
{\cal L}_{aq} &=&{\cal L}_{qa} = -T^{2}\left[\mu + \sqrt{\mu^{2} - 4J^{2}}\right] z_T, \label{Eq:OnsagerDiffusive}
\end{eqnarray}
where we omitted the $k$-dependence of the relaxation distance $z_T$. Finally, within the linear response theory ($\nabla T\sim\Delta 
T/N \ll \bar{T}\equiv (T_{L} + T_{R})/2$, $\nabla\mu\sim\Delta \mu/N \ll \bar{\mu}\equiv (\mu_{L} + \mu_{R})/2$), we neglect 
the $x$-dependence of temperature $T(x)$ and chemical potential $\mu(x)$ and approximate the Onsager elements as ${\cal L}_{ij}
(T(x), \mu(x))\approx {\cal L}_{ij}(\bar{T}, \bar{\mu})$. We find that, contrary to the ballistic domain, $j_a,j_q \propto 1/N$ featuring 
the so-called normal transport, where the transport coefficients $\kappa, \sigma$ are independent on the system size and Fourier's law holds.

{\it One-parameter scaling theory --} Next we have established a one-parameter scaling theory that controls the variation 
of $j_a$ as the size of the junction increases. Specifically, we find that the rescaled power current $p_N (\chi,J,\bar{T},
\bar{\mu})$ through a junction of length $N$ which is attached to two reservoirs with mean temperature 
and chemical potential $\bar{T}, \bar\mu$ obeys a one-parameter scaling, i.e.,
\begin{equation}\label{Eq:scaling}
 \frac{\partial p_N}{\partial ln N}=\beta(p_N),
\end{equation}
where $\beta$ is a function of $p_N$ alone. The rescaled power current is $p_N(\chi,J,T,\mu)\equiv \frac{j_{a}}{j_{a}^{(\chi)}}$ where 
$j_{a}^{(\chi)}$ is the power current in the ballistic regime.  

The scaling ansatz of \cref{Eq:scaling} is equivalent to postulating the existence of a function $f(x)$ such that
\begin{equation}\label{Eq:scaling2}
p_N=f(\lambda\equiv N/l_T)\propto \left\{
\begin{array}{cc}
1;&\quad \lambda \ll {\cal O}(1) \\
\frac{1}{N};&\quad \lambda \gg {\cal O}(1)
\end{array}
\right..
\end{equation}
The scaling parameter $\lambda\equiv N/l_T$, encodes all the information about the relaxation process towards a 
non-equilibrium (quasi-)steady state current and describes the number of thermalized segments with length $l_T
\sim v_g z_T$ contained in a junction of length $N$. In the ballistic limit $l_T\gg N$, the junction consists of a single 
segment and the nonlinear interactions are unable to enforce thermalization of the modes. Therefore, the transport 
is essentially ballistic and $\frac{j_a}{j_a^{(\chi)}}\approx 1$. On the other hand, when $N\gg l_T$, the network 
consists of a number of $\lambda\gg 1$ uncorrelated segments. This situation is reminiscent of the law of additive 
resistances connected in series. As in this case, the total current decays inversely proportional to the number of segments 
$j_a\propto 1/N$. The relaxation distance that defines $l_T$ is $z_T\propto \frac{\chi^{2}a^{2}}{J}\tanh\left[\frac{T}
{\zeta J a}\right]$ and has been previously evaluated in Ref. \onlinecite{Ramos2023} ($a\sim 1$ is the average value 
of norm per site, and $\zeta\approx 8$ is a best fitting parameter).

\begin{figure}
\center\includegraphics[width=1\columnwidth,keepaspectratio]{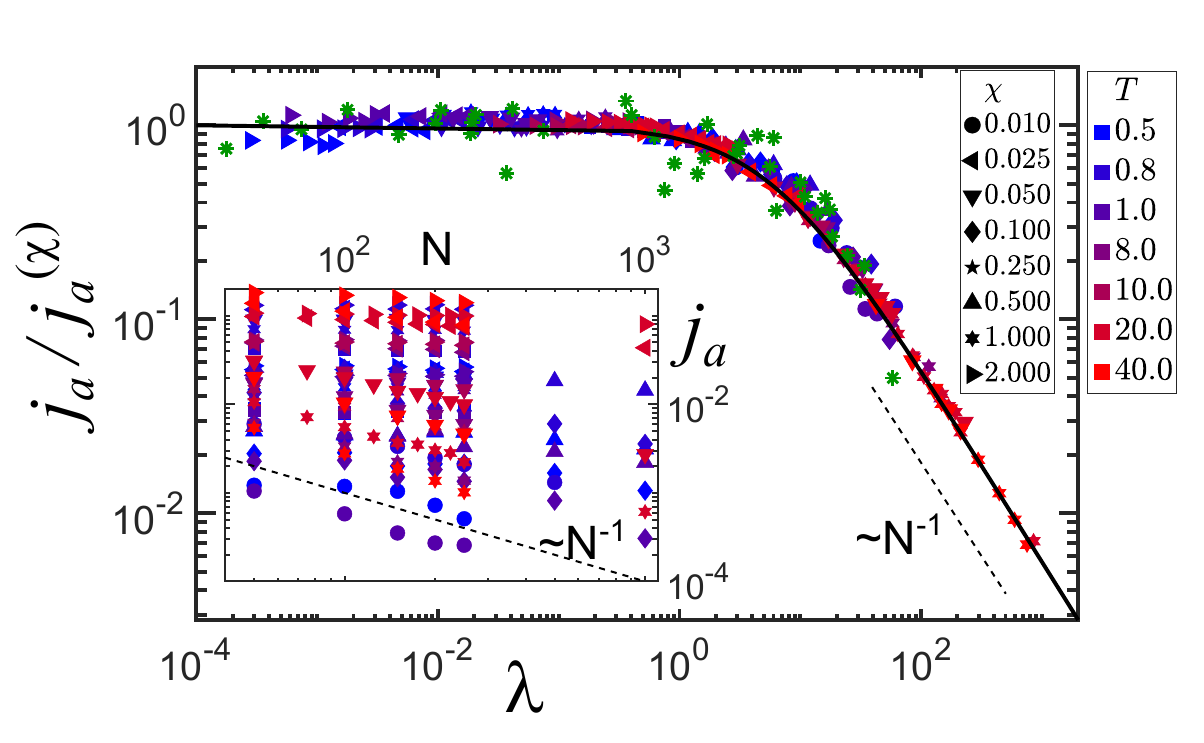}
\caption{
Normalized power current $j_a/j_a^{(\chi)}$ versus the scaling parameter $\lambda = N/l_T$ for a 1D junction,
of (transverse) size $N$. The different values of the current $j_a$ for a variety of parameters $(\chi,J,\bar{T},\bar{\mu})$
are shown in the inset. The black line is the interpolating function \cref{Eq:scalingFunc}, while the dashed line indicates 
a $1/N$ behavior and is drawn to guide the eye.
%
}
\label{fig2}
\end{figure}

To validate our scaling ansatz \cref{Eq:scaling}, we have performed detailed numerical simulations for a variety of nonlinear 
coefficients $\chi$ and system sizes $N$ for both high and low temperatures $T$. Our results are summarized in Fig. \ref{fig2}, 
where we report the outcome of our simulations using two methods: (1) Modeling the large collections of $M$ modes in the 
bundles by Monte-Carlo reservoirs \cite{Iubini2012} with effective thermostats at fixed $(T,\mu)$ (see filled symbols in Fig.  
\ref{fig2}). (2) Solving numerically the TCMT \cref{Eq:TCMT} for the whole system (reservoirs + junction) with reservoirs 
consisting of $M=1600$ coupled modes forming a square lattice (see Supplementary Material). A possible interpolating law that describes 
our data (including the crossover regime) is 
\begin{equation}
 f(\lambda)=\frac{1}{1+\lambda/\lambda_{\ast} },
 \label{Eq:scalingFunc}
\end{equation}
where $\lambda_{\ast} = 0.18$ is the best fitting parameter. 

{\it Photonic Wiedemann-Franz law and thermal current --} In thermoelectric devices, the Wiedemann-Franz (WF) law 
connects the thermal and current conductivity in a simple manner: their ratio is proportional to the temperature i.e. $\varkappa/
\sigma = L T$ with a proportionality constant $L$ which, in typical metals, takes a universal value \cite{AshcroftMermin}.  
This proportionality relation, essentially describes the fact that a good electrical conductor, is also an efficient heat conductor, 
with thermal to electric conductivity ratio proportional to temperature -- a property that is rooted on the fact that heat and 
charge currents are associated to the flux of the same (quasi)particles. Deviations from the WF law signify the existence 
of multiple transport mechanisms for thermal and/or electrical fluxes, that ultimately allow for the independent control of 
electrical and thermal transport \cite{Craven2020,Benenti2020}.

\begin{figure}
\center
\includegraphics[scale = 0.3]{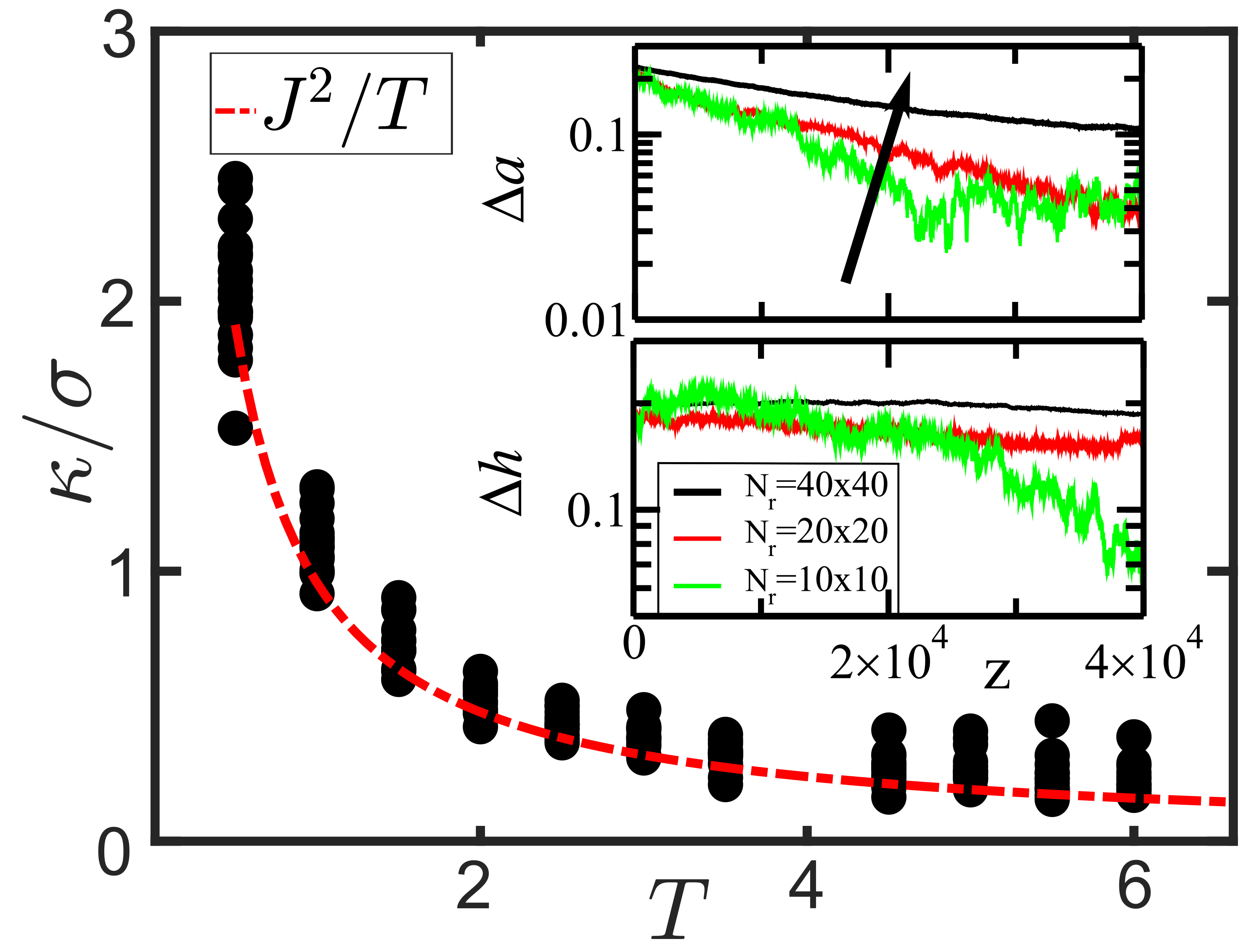}\\
\caption{The ratio of thermal to power conductivity $\varkappa/\sigma$ versus temperature (black dots) 
for different values of nonlinearities $\chi =$ 0, 0.025, 0.05, 0.1, 0.25, 0.5, 0.75, 1.0,  and junction sizes: 
$N = $ 150, 300, 600. The theoretical prediction \cref{Eq:WF} (red dashed line) indicates a $\propto1/T$ 
behavior, signifying a separation of relaxation processes between thermal and power transport. In the insets 
we show the difference of power (upper inset) and energy (lower inset) densities between the left and right 
reservoirs for three different reservoir sizes $M$. As $M$ increases the two reservoirs maintain for longer 
time their initial energy and power densities allowing us to extract the (quasi)-steady-state values for the 
thermal and power currents.}
\label{fig3}
\end{figure}

The equivalent of charge (particle) conductivity in photonics, is power conductivity. It is natural, therefore, to extend the 
above definition of WF law and analyze the corresponding ratio of thermal conductivity to power conductivity. By 
combining Eqs. (\ref{Eq:OnsagerBallistic},\ref{Eq:OnsagerDiffusive}) we obtain  
\begin{equation}\label{Eq:WF}
\frac{\varkappa}{\sigma} \approx \frac{J^{2}}{T}, \,{\rm for}\,\, |\mu|\gg (2J),
\end{equation}
signifying a novel form of WF law which occurs in MMNPC. This theoretical prediction is nicely confirmed by our 
simulations using Monte-Carlo optical reservoirs (see main panel of  \cref{fig3}). The various values of the nonlinear 
coefficient $\chi$ and junction size $N$ that have been used in these simulations were chosen to guarantee that we have 
spanned the full transport domain from ballistic to diffusive regimes. 

The inverse temperature dependence of the ratio $\varkappa/\sigma$ signifies the decoupling of thermal and power 
transfer. A similar phenomenon 
has been also observed in ultracold atomic gases \cite{FHM16} implying that atom-atom interactions affect the associated 
thermal and particle conductivities in radically different ways. Specifically, it was shown that there is a time-scale separation 
for the equilibration of temperature and particle imbalances between the two reservoirs. We have demonstrated the different 
equilibration times by performing dynamical simulations with small composite (junction + reservoirs) system sizes. In 
these simulations we have utilized a microcanonical approach for the whole system and found different relaxation scales 
for the internal energy and power differences between the two reservoirs (see insets of \cref{fig3}).

Let us finally point out that unlike the familiar case of metals, in the developed optical kinetics framework, the 
results for the WF law are sensitive to the definition of $\kappa$. If, for example, we had defined the thermal 
conductivity under the constraint $\nabla \mu=0$ (as opposed to the traditional $j_a=0$ used above), we will 
end up with a different result for $\kappa/\sigma$ (see Eqs. (S12,S19) of the Supplementary material). In this 
case, at the high temperature limit where $n^{(0)}\approx 1$, we get the familiar expression $\kappa/\sigma=T$.

{\it Conclusion --} We have developed an optical kinetics framework that can be utilized to predict the complex response 
of nonlinear heavily multimoded optical junctions when they are brought in contact with optical reservoirs. The transport 
coefficients derived here can be utilized for the design of all-optical cooling protocols. It will be interesting 
to extend our formalism to include localization effects due to the presence of transverse disorder or the influence of paraxial 
noise in the steady-state currents.

%

\begin{acknowledgments}
We acknowledge partial support from Simons Foundation for Collaboration in MPS grant No. 733698 
\end{acknowledgments}



\newpage
   .
\newpage
\onecolumngrid
\begin{center}
{\huge Supplementary Material}
\end{center}

\setcounter{equation}{0}
\setcounter{figure}{0}
\setcounter{table}{0}
\setcounter{page}{1}
\makeatletter
\renewcommand{\theequation}{S\arabic{equation}}
\renewcommand{\thefigure}{S\arabic{figure}}
\renewcommand{\bibnumfmt}[1]{[S#1]}
\renewcommand{\citenumfont}[1]{S#1}

\section{Coupled Mode Theory Modeling}

\subsection{Equations of motion.-}
The beam dynamics can be described by a coupled mode theory (CMT)\cite{SRamos2020}
\begin{equation}\label{Eq:SupTCMT}
i\frac{d\psi_{l}}{dz} =  - \sum_{j} J_{lj} \psi_j + \chi|\psi_{l}|^{2} \psi_l,
\end{equation}
where the propagation through the $z$-direction plays the role of time in a DNLSE. Here, 
$\psi_l$ is the complex field amplitude at the $l$-th node of the network, $\eps_{l} =- J_{ll}$ represent the optical 
on-site potentials (e.g. propagation constants) in the absence of coupling or nonlinearities, and the last term dictates 
the non-linearity due, for instance, to a Kerr effect.  The coupling coefficients $J_{lj}^{} = J_{jl}^{\ast}$ between sites
$l, j$ dictate the connectivity of the photonic network. For the junction, $J_{lj} = J\delta_{l, j\pm 1}$, and 
$\epsilon_{l} = \epsilon = 0$.

The field dynamics given by Eq. \ref{Eq:SupTCMT} ensures that there are two constants of motion that represent the internal 
energy and the total optical power, respectively,
\begin{eqnarray}\label{Eq:SupHamiltonianGeneral}
\mathcal{H}(\{\vc{\psi}\}) &=& -\sum\limits_{i\neq l}J_{lj}\psi_{l}^{\ast}\psi_{j} +
\sum\limits_{l}\left[\eps_{l} + \frac{\chi}{2}|\psi_{l}|^{2}\right]|\psi_{l}|^{2}\approx \sum_\alpha \omega_\alpha |C_\alpha|^2\equiv E,\nonumber
\\
\mathcal{N}(\{\vc{\psi}\}) &=& \sum\limits_{l}\left|\psi_{l}\right|^{2}= \sum_\alpha |C_\alpha|^2 \equiv A.
\end{eqnarray}
Like in the main text, $C_\alpha$ represents the projection coefficient to the supermode $\alpha$.
To arrive to the rhs of Eq. (\ref{Eq:SupHamiltonianGeneral}) we have assumed weak nonlinearity. 
The presence of  nonlinearities provides a mechanism for mode-mixing processes that lead, ultimately, to thermal equilibrium. 
If the nonlinearity is weak, such a thermal state corresponds to a supermode occupation 
$n (\omega_{\alpha}) = \left\langle|C_{\alpha} |^2\right\rangle= \frac{T}{\omega_{\alpha} -\mu}$, i.e., given by the Rayleigh-Jeans (RJ)
statistics \cite{SRamos2020,SShi2021, SWu2019}, where $\{\omega_{\alpha}\}$ represents the linear spectrum, and the Lagrange multipliers
$T = T(E, A)$, $\mu = \mu(E, A)$ correspond to the optical temperature and chemical potential.

\subsection{CMT simulations.-}

For our CMT simulations, we have employed two methods: 
$(1)$ We conduct 'time' simulations by integrating \cref{Eq:SupTCMT} for the entire system, encompassing the junction and the reservoirs.
$(2)$ Alternatively, we adopt the Monte-Carlo thermostat approach \cite{SIubini2012}, where we replace the extensive collections of modes 
within the reservoirs with effective thermostats that enforce specific temperature and chemical potential values $(T,\mu)$. 
Full simulations are time-consuming, rendering the second method, reliant on Monte Carlo reservoirs, highly desirable for 
our calculations. Both methods are elaborated upon below.

\subsubsection{Full CMT simulations.-}
Our CMT simulations have been performed by integrating \cref{Eq:SupTCMT} using an 8-th order three-part split symplectic integrator 
algorithm \cite{SGMS96,SCN96,SRamos2020,SShi2021}. Such an algorithm guarantees the conservation of both the total optical power $A$ and 
internal energy $E$ during the whole evolution with a numerical accuracy ensuring conservation up to ${\cal O}(10^{-8})$. 

The lattice connectivity is such that the reservoirs consist of $N_r=40\times 40=1600$ sites forming a square lattice with
coordination number $z=2$ and a uniform nearest-neighbor coupling constant $J_r=1$. To prevent spectral degeneracies, we set 
the optical on-site potentials to be random numbers drawn from a uniform distribution within the 
range $[-\frac{W}{2},\frac{W}{2}]$ with $W=0.5$. We have verified that a reservoir size $N_r=1600$ guarantees a 
(quasi-)steady state of the currents.

On the other hand, the junction corresponds to a 1D chain with coordination number $z_{0}=1$, and it is connected to the reservoirs
at the corners with a junction-reservoir coupling $J_{j-r}=0.2 J$.

In our simulations, we have prepared the initial states of the reservoir $L(R)$ at energy and optical power 
$(E_{L(R)},A_{L(R)})$, that after thermalization, will converge toward a RJ distribution with optical temperature 
and chemical potential $(T_{L(R)},\mu_{L(R)})$. Following this protocol, we have generated an ensemble of $M\gtrsim 50$
realizations by setting random phases for the supermode complex amplitudes $C_{\alpha}$ and integrated \cref{Eq:SupTCMT} for 
times $t\gtrsim 5\times 10^4$ in units of $J^{-1}$. 

\subsubsection{Monte Carlo thermostats.-}
This method, inspired in Ref. \onlinecite{SIubini2012}, enables us to replace the large number of modes in the reservoirs with an effective thermostat, allowing us to concentrate
solely on the field dynamics through the junction. In turn, this results in more efficient simulations. Then, in the present study, 
the majority of our numerical results were obtained using this approach.

At the Monte-Carlo thermostat, the field amplitudes can undergo a random perturbation (at random times) 
$\psi_m\to \psi_m+\delta\psi_m$ at sites $m$ connected to the reservoirs. Such perturbations are accepted or denied based on the standard Metropolis algorithm, which evaluates the cost function 
 \[
 \xi(\delta \psi_m ) = \exp\left\{-\frac{\Delta \mathcal{H} - \mu \Delta \mathcal{N}}{T} \right\},
 \]
where $T$, $\mu$ are temperature and chemical potential of the corresponding thermostat. $\Delta \mathcal{H}$ and $\Delta \mathcal{N}$
correspond to the variation of the energy and optical power of the chain due to the perturbation, i.e., 
$\Delta \mathcal{H}\equiv \mathcal{H}(\vc{\psi+\delta\psi}^{})-\mathcal{H}(\vc{\psi}^{})$ 
and  $\Delta \mathcal{N}\equiv \mathcal{N}(\vc{\psi+\delta\psi}^{})-\mathcal{N}(\vc{\psi}^{})$.

\subsubsection{Evaluation of the currents.-}
After neglecting transients, we can extract the optical power current $j_a$ and the internal energy current $j_h$ 
through the junction via  
\begin{equation}\label{Eq:SupCurrentsMicroscopic}
\begin{aligned}
&j_a \equiv \langle j_{a,n} \rangle_{n,z} = i J\left\langle \psi_n \psi^{\ast}_{n-1} - \psi_{n-1}\psi^{\ast}_{n}\right\rangle_z,\\
&j_h \equiv \langle j_{h,n} \rangle_{n,z} =  J\left\langle\dot{\psi}_n \psi_{n-1}^{\ast} + \psi_{n-1}\dot{\psi}_n^{\ast}\right\rangle_z,
\end{aligned}
\end{equation} 
which follow from the continuity equations $\frac{d a}{dz} + \frac {\partial j_{a}}{\partial x} = 0$, and $\frac{d h}{dz} +
\frac {\partial j_{h}}{\partial x} = 0$, respectively\cite{SIubini2012}. The values of $j_a$ and $j_h$ do not depend on the position,
and an average over different sites $n$ and on 'time' $z$ help in reducing fluctuations. However, an alternative approach consists in
extracting the currents by monitoring the variations of the optical power and internal energy at the reservoirs.
We have numerically verified that both approaches give the same results, nevertheless, the latter is more efficient in
reducing the number of realizations in the average. These methodologies can be used with both the full CMT simulations or the
Monte Carlo thermostats.


\section{Linear Response Theory\label{Sec:Phenomenology}}
In the (quasi-)stationary regime, the power density $a\equiv\mathcal{N}/N$ and energy density $h\equiv\mathcal{H}/N$ at any segment
$dx$ inside the junction (of length $N\gg dx$) satisfy the continuity equations\cite{SCallen1985,SSaito2010} 
\begin{equation}
\frac{d a}{dz} + \nabla j_{a}  = 0,\quad
\frac{d h}{dz} + \nabla j_{h}  = 0,
\end{equation}
where we assumed that the segment $dx$ includes many unit cells while $a(x)$ and $h(x)$ are the values of these densities at a position
(site) $x$. Assuming that each segment $(x,x+dx)$ is at local equilibrium, we can define a local
temperature and chemical potential, $T=T(x)$, $\mu=\mu(x)$, and thus study the transport using the framework of linear response
theory following the Onsager matrix formalism \cite{SCallen1985,SSaito2010}:
\begin{equation}\label{Eq:SupOnsagerPrime}
\begin{aligned}
j_{a} = \mathcal{L}^{\prime}_{aa}\nabla\left(-\frac{\mu}{T}\right) + \mathcal{L}^{\prime}_{ah}\nabla\left(\frac{1}{T}\right),\\
j_{h} = \mathcal{L}^{\prime}_{ha}\nabla\left(-\frac{\mu}{T}\right) + \mathcal{L}^{\prime}_{hh}\nabla\left(\frac{1}{T}\right),
\end{aligned}
\end{equation} 
where $\nabla (-\mu/T)$, $\nabla (1/T)$ are so-called thermodynamical forces or affinities. An alternative formulation 
of the Onsager matrix results from introducing the heat current $j_{q} = j_{h} - \mu j_{a}$, 
\begin{equation}\label{Eq:SupOnsager}
\begin{aligned}
j_{a} = \mathcal{L}_{aa}\left(\frac{-1}{T} \nabla\mu\right) + \mathcal{L}_{aq}\nabla\left(\frac{1}{T}\right),\\
j_{q} = \mathcal{L}_{qa}\left(\frac{-1}{T} \nabla\mu\right) + \mathcal{L}_{qq}\nabla\left(\frac{1}{T}\right), 
\end{aligned}
\end{equation} 
where now the new affinities read $\nabla (-\mu)/T$, $\nabla (1/T)$. Both formulations are equivalent, thus, the relation
among the coefficients of the Onsager matrices $\mathcal{L}$ and $\mathcal{L}^\prime$ are given by 
\begin{equation}\label{Eq:SupOnsagerL}
\begin{aligned}
&\mathcal{L}_{aa} = \mathcal{L}^{\prime}_{aa}, \quad \mathcal{L}_{aq} = \mathcal{L}^{\prime}_{ah} - \mu \mathcal{L}^{\prime}_{aa},\\
&\mathcal{L}_{qa} = \mathcal{L}^{\prime}_{ha} - \mu \mathcal{L}^{\prime}_{aa}, \quad \mathcal{L}_{qq} = \mu^{2}\mathcal{L}^{\prime}_{aa} + \mathcal{L}^{\prime}_{hh} - 2\mu \mathcal{L}^{\prime}_{ah},
\end{aligned}
\end{equation}
or 
\begin{equation}\label{Eq:SupOnsagerLprime}
\begin{aligned}
&\mathcal{L}^{\prime}_{aa} = \mathcal{L}_{aa}, \quad \mathcal{L}^{\prime}_{ah} =  \mathcal{L}_{aq} + \mu \mathcal{L}_{aa},\\
&\mathcal{L}^{\prime}_{ha} = \mathcal{L}_{qa} + \mu \mathcal{L}_{aa}, \quad \mathcal{L}^{\prime}_{hh} = \mathcal{L}_{qq}  + 2\mu \mathcal{L}_{aq}  + \mu^{2}\mathcal{L}_{aa}.
\end{aligned}
\end{equation}
 As it follows from the Onsager theorem, $\det\Matr{L} = \det\Matr{L}^{\prime}\geqslant 0$, $\mathcal{L}_{aa}\geqslant 0$, 
 $\mathcal{L}_{qq}\geqslant 0$, $\mathcal{L}_{aa}^{\prime}\geqslant 0$, $\mathcal{L}_{hh}^{\prime}\geqslant 0$, 
 $\mathcal{L}_{aq} = \mathcal{L}_{qa}$, $\mathcal{L}_{ah}^{\prime} = \mathcal{L}_{ha}^{\prime}$. 
 The latter are the well-known Onsager reciprocity relations.

 All coupled transport properties are characterized by the Onsager matrix or, equivalently, by the transport coefficients,
 which include the optical power conductivity, $\sigma$, thermal conductivity, $\varkappa$, and the Seebeck coefficient, $S$,
 defined as follows:
\begin{subequations}\label{Eq:SupTransportCoeffs}
\begin{equation}
j_{a} = -\sigma \nabla \mu\Big|_{\nabla T = 0} \Longrightarrow \sigma =  \frac{\mathcal{L}_{aa}}{T},
\end{equation}
\begin{equation}
j_{q} = -\varkappa \nabla T\Big|_{j_{a} = 0}\Longrightarrow \varkappa = \frac{\det\Matr{L}}{T^{2}\mathcal{L}_{aa}},
\end{equation}
\begin{equation}
\nabla \mu_{\rm ind} = -S \nabla T\Big|_{j_{a} = 0}\Longrightarrow S = \frac{1}{T}\frac{\mathcal{L}_{aq}}{\mathcal{L}_{aa}},
\end{equation}
\end{subequations}  
where $\nabla \mu_{\rm ind}$ stands for the chemical potential gradient induced by a temperature gradient without optical power
exchange between the thermostats.


\section{Ballistic transport and Landauer theory \label{Sec:Landauer}}

In the linear limiting case $\chi \approx 0$, there are no significant mode-mixing interactions  and 
the modes of the junction are freely propagating waves. Thus, the transport is essentially ballistic and currents can be analyzed using
the linear scattering framework provided by Landauer's theory \cite{SImry1999},
\begin{equation}\label{Eq:SupCurrentsLandauer}
\begin{aligned}
&j_a = \int \,d\omega \cdot t(\omega)\cdot\left[n_{\rm L}(\omega, \mu_{\rm L}, T_{\rm L}) - n_{\rm R}(\omega, \mu_{\rm R}, T_{\rm R})\right],\\
&j_h = \int \,d\omega \cdot t(\omega)\cdot \omega\cdot\left[n_{\rm L}(\omega, \mu_{\rm L}, T_{\rm L}) - n_{\rm R}(\omega, \mu_{\rm R}, T_{\rm R})\right],
\end{aligned}
\end{equation} 
where $n_{\rm L,R}(\omega, \mu_{\rm L,R}, T_{\rm L,R})\equiv T_{\rm L,R} /(\omega -\mu_{\rm L,R})$ are the equilibrium distribution
functions of the reservoirs. The factor $t(\omega)$ corresponds to the transmittance through the junction, which in our case 
$t(\omega)\approx t_{0}$ for $\omega\in [-2J,2J]$, and $t(\omega)=0$ otherwise. Then, by evaluating the integrals in \cref{Eq:SupCurrentsLandauer}
\begin{equation}
\begin{aligned}
j_a &= t_{0}\Bigg[T_{\rm L}\ln\left(\frac{-\mu_{\rm L} + 2J}{-\mu_{\rm L} - 2 J}\right) - T_{\rm R}\ln\left(\frac{-\mu_{\rm R} + 2 J}{-\mu_{\rm R} - 2J}\right)\Bigg],\\
j_h &= t_{0}\Bigg[4J(T_{\rm L} - T_{\rm R}) + 
  \mu_{\rm L}T_{\rm L}\ln\left(\frac{-\mu_{\rm L} + 2J}{-\mu_{\rm L} - 2 J}\right)
  - \mu_{\rm R}T_{\rm R}\ln\left(\frac{-\mu_{\rm R} + 2 J}{-\mu_{\rm R} - 2J}\right)\Bigg].
\end{aligned}
\end{equation}
The equations above are valid for arbitrary values of the temperature and chemical potential of the reservoirs. 
Nevertheless, in order to obtain the Onsager coefficients, it is useful to consider the linear response regime, where $|\Delta\mu|\ll -\mu - 2J$ and $\Delta T\ll T$. Then,
\begin{equation}\label{Eq:SupBallisticCurrents}
\begin{aligned}
&j_{a} = -t_{0}\left[\frac{4JT}{\mu^{2} - 4J^{2}}\Delta\mu +\ln\left(\frac{-\mu + 2J}{-\mu -2J}\right)\Delta T\right],\\
&j_{h} = -t_{0}\left[4J\Delta T + T\ln\left(\frac{-\mu + 2J}{-\mu -2J}\right)\Delta\mu\right] + \mu j_{a},\\
&j_{q} = -t_{0}\left[4J\Delta T + T\ln\left(\frac{-\mu + 2J}{-\mu -2J}\right)\Delta\mu\right],
\end{aligned}
\end{equation}
where we introduced $T \equiv (T_{L} + T_{R})/2$, $\mu \equiv (\mu_{L} + \mu_{R})/2$, $\Delta T \equiv T_{R} - T_{L}$, 
$\Delta\mu \equiv \mu_{R} - \mu_{L}$, and we omitted terms $\mathcal{O}[(\Delta\mu/\mu)^{2}]$,   $\mathcal{O}[(\Delta T/T)^{2}]$, 
$\mathcal{O}[(\Delta\mu/\mu)(\Delta T/T)]$. 

With the analytical expressions obtained above, it is straightforward to obtain the Onsager matrix elements
\begin{equation}\label{Eq:SupOnsagerBallistic}
\mathcal{L}_{aa} = t_{0}N\frac{4J T^2}{\mu^2 - 4J^{2}},\quad 
\mathcal{L}_{aq} = \mathcal{L}_{qa} = -t_{0}NT^{2}\ln\left(1 - \frac{4J}{-\mu + 2J}\right), \quad
\mathcal{L}_{qq} = t_{0}N4JT^{2},
\end{equation}
where we have replaced the gradients $\nabla (\cdot) \longleftrightarrow \Delta(\cdot)/N$.
At the same time, we can write the transport coefficients in the ballistic regime as
\begin{equation}\label{Eq:SupTransportCoeffsBallistic}
\begin{aligned}
&\sigma = N\frac{4J T}{\mu^2 - 4J^{2}}t_0,\\
&\varkappa  = 4JN\left[1 - \frac{\mu^2 - 4J^2}{16J^2}\ln^2\left(1 - \frac{4J}{-\mu + 2J}\right)\right]t_0,\\
&S = \frac{\mu^2 - 4J^2}{4JT}\ln\left[\frac{-\mu - 2J}{-\mu+2J}\right].
\end{aligned}
\end{equation}
The distinctive feature of the ballistic regime, as confirmed by \cref{Eq:SupBallisticCurrents} and \cref{Eq:SupTransportCoeffsBallistic}, is that 
currents remain independent of the system size (N), while the conductivity scales linearly with N. In the ballistic regime, all modes 
are delocalized, and no scattering mechanisms are involved. Transport in this regime can be visualized as a single direct transfer from 
one reservoir to the other. 

\subsection{Effect of a weak nonlinearity}

Next, we consider the impact of weak nonlinearity in the junction on currents and transport coefficients. As discussed in the main text,
the Landauer approach (Eq. \ref{Eq:SupCurrentsLandauer}) remains applicable, with the modification of introducing a transmittance that 
maintains a constant value, denoted as $t_0$ within a shifted frequency window defined by $\omega \in [-2J+2\chi a, 2J+2\chi a]$. This 
transmittance is zero everywhere outside this window. Here, we provide an explanation of the mechanisms responsible for this frequency 
shift.

While a weak nonlinearity cannot enforce sufficient mode-mode mixing, it can induce a nonlinear
frequency shift on the $\alpha$-th supermode frequency \cite{SBasko2014}
\begin{equation}
\label{eq:Supshifted_frequency}
\tilde{\omega}_{\alpha} ^{(J)} \approx \omega_{\alpha}^{(J)} + 2 a\chi.
\end{equation}
Such a shift can be seen by considering
the Hamiltonian of the junction (in absence of the reservoirs) in the normal modes representation
\begin{equation}
\mathcal{H} = \sum\limits_{\alpha}\omega_{\alpha}^{(J)}|C_{\alpha}|^{2} + \frac{\chi}{2}\sum\limits_{\alpha\beta\gamma\delta}V_{\alpha\beta\gamma\delta}C_{\alpha}^{\ast} C_{\beta}^{\ast} C_{\gamma}^{} C_{\delta}^{}, 
\end{equation}
where the mode mixing factor
\begin{equation}
V_{\alpha\beta\gamma\delta} = \sum\limits_{n}\phi^{\ast}_{\alpha}(n)\phi^{\ast}_{\beta}(n) \phi_{\gamma}(n)\phi_{\delta}(n),
\end{equation}
describes the strength of the mode-mode interactions associated with the nonlinear mixing between supermodes.
There, $\phi_{\alpha}(n)\equiv \langle n| \phi_\alpha\rangle$ represents a projection of the onsite amplitudes on the normal modes.
The so-called \textit{secular} terms, for which either $\alpha = \delta$, $\gamma = \beta$, or $\alpha = \gamma$, $\beta = \delta$,
are the responsible to the frequency shift by contributing to the integrable Hamiltonian component
\begin{equation}
\mathcal{H}_0  = \sum\limits_{\alpha}|\phi_{\alpha}|^{2}\left[\omega_{\alpha}^{(J)} + 2\chi\sum\limits_{\beta}V_{\alpha\beta\beta\alpha}|\phi_{\beta}|^{2}\right],
\end{equation} 
while the remaining terms, $\mathcal{H}-\mathcal{H}_0$, representing the integrability breaking processes, produce mode-mixing  interactions. 
By considering that $\langle |C_\alpha|^2\rangle \approx a$ and invoking to the orthogonality of the supermodes $\phi_\alpha$, we arrive to \cref{eq:Supshifted_frequency}.

As a result, the frequency correction \cref{eq:Supshifted_frequency} produces a uniform shift of the dispersion relation through 
the junction that, in turn, results in wave propagation with a constant transmittance $t_0$ within the frequency band described above.

\section{Kinetic equation and diffusive transport\label{Sec:Diffusion}}

In this section, we evaluate the integrals in Eq. (8) of the main text to obtain the currents and Onsager coefficients.
To that purpose, we introduce the linearized Kinetic Equation Eq. (7) of the main text into Eq. (8) and we obtain
\begin{equation}
\begin{aligned}
&j_{a} = 
-\frac{\nabla\mu}{T}  \int\limits_{-\pi}^{+\pi}  \frac{d k }{2\pi} \, v_g^{2}(k)\, \left[n^{(0)}(k)\right]^{2}\, z_T(k) +
\nabla \left(\frac{1}{T}\right)\int\limits_{-\pi}^{+\pi}  \frac{d k }{2\pi} T \,v_g^{2}(k)\,n^{(0)}(k)\,z_T(k),\\
&j_{q} = -\frac{\nabla\mu}{T}  \int\limits_{-\pi}^{+\pi}  \frac{d k }{2\pi} T \,v_g^{2}(k)\,n^{(0)}(k)\,z_T(k) +
\nabla\left(\frac{1}{T}\right)\int\limits_{-\pi}^{+\pi}  \frac{d k }{2\pi} T^{2}\, v_g^{2}(k)\,z_T(k).
\end{aligned}
\end{equation}
The relaxation length $z_T(k)$ generally depends on $k$. Nevertheless, here we move forward by relying on a rough approximation where 
we assume $z_T(k)\approx z_T=\mathrm{cons.}$, and the order of magnitude estimate of $z_T\propto \frac{\chi^{2}a^{2}}
{J}\tanh\left[\frac{T}{\zeta J a}\right]$ has been previously evaluated in Ref. \onlinecite{SRamos2023}
($a=A/N\sim 1$ is the average value of norm per site inside the junction, and $\zeta\approx 8$ is a best fitting parameter).
Using $\omega_\alpha^{(J)} = -2J\cos( k_\alpha)$, we derive the Onsager matrix elements,
 \begin{equation}\label{Eq:SupOnsagerDiffusive}
 \begin{aligned}
 &\mathcal{L}_{aa} = 4J^{2}T^{2}\, z_T \frac{1}{\pi} \int\limits_{0}^{\pi}dk \frac{\sin^{2} k}{\left[-2J\cos k - \mu\right]^{2}}  
 = -T^{2}\left[1 + \frac{\mu}{\sqrt{\mu^{2} - 4J^{2}}}\right]\, z_T ,\\
 & \mathcal{L}_{aq} = \mathcal{L}_{qa} = 4J^{2}T^{2} z_T \frac{1}{\pi}\int \limits_{0}^{\pi} dk\frac{\sin^{2} k} {-2J\cos k  - \mu} 
 = -T^{2}\left[\mu + \sqrt{\mu^{2} - 4J^{2}}\right]z_T ,\\
 &\mathcal{L}_{qq} = 4J^{2}T^{2}z_T \frac{1}{\pi}\int\limits_{0}^{\pi}dk\sin^{2} k  = 2 J^{2}T^{2} z_T .
 \end{aligned}
 \end{equation}
 
 Finally, the transport coefficients in the diffusive regime read
 \begin{equation}\label{Eq:SupTransportCoeffsDiffusive}
\begin{aligned}
&\sigma = T\left[\frac{1}{\sqrt{1 - \left(\frac{2J}{\mu}\right)^{2}}}-1\right]z_T,\\
&\varkappa = \mu^{2}\left[1 - \frac{1}{2}\left(\frac{2J}{\mu}\right)^{2} -\sqrt{1 - \left(\frac{2J}{\mu}\right)^{2}}\right]z_T,\\
&S = \frac{\sqrt{\mu^{2} - 4J^{2}}}{T}.
\end{aligned}
\end{equation}


\section{Wiedemann-Franz Law}

Here, we provide analytical expressions for the Wiedemann-Franz Law (WFL) in both the ballistic and diffusive regime. 
For this purpose, we invoke the results that we have obtained in the previous sections, and the WFL reads
\begin{equation}\label{Eq:SupWF}
\frac{\varkappa}{\sigma} = 
\left\{
\begin{aligned}
&\left(\mu^2 - 4J^{2}\right)\left[1 - \frac{\mu^2 - 4J^2}{16J^2}\ln^2\left(1 - \frac{4J}{-\mu + 2J}\right)\right]\frac{1}{T}
\xrightarrow {|\mu|\gg2J} \frac{J^2}{T} \quad (\text{ballistic}),\\
&\frac{\mu^{2}}{2}\left[\sqrt{1 - \left(\frac{2J}{\mu}\right)^{2}} - \left[1 - \frac{1}{2}\left(\frac{2J}{\mu}\right)^{2}\right] + \frac{1}{2}\left(\frac{2J}{\mu}\right)^{2}\right]\frac{1}{T}\xrightarrow {|\mu|\gg2J} \frac{J^2}{T}\quad (\text{diffusive}).
\end{aligned}
\right.
\end{equation}
Notice that this ratio does not depend on $t_0$ nor $z_T$.
In both the ballistic or the diffusive regime when $|\mu|\gg 2J$, the WFL takes the asymptotic form
$\frac{\varkappa}{\sigma}\approx \frac{J^2}{T}$ that corresponds to the Eq. (12) discussed in the main text.

To confirm this result numerically, we calculate the ratio $\varkappa/\sigma$ from the elements of the Onsager matrix. The latter 
can be numerically obtained by extracting the energy and power currents under appropriate preparations of the corresponding
affinities, as we describe next.
First, we consider the situation where the temperature of the reservoirs are the same, $T_L=T_R$, resulting in a vanishing thermal force
$\nabla (1/T)=0$. We denote the associated currents as $j_a^{(T)}=\mathcal{L}^\prime_{aa}\cdot \nabla \left( \frac{-\mu}{T}\right)$ and 
$j_h^{(T)}=\mathcal{L}^\prime_{ha}\cdot \nabla \left( \frac{-\mu}{T}\right)$, which are extracted from various
numerical simulations (with $J = 1$) at different temperature values ranging from $T=0.5$ to $T=6$ and chemical potential with mean 
value $\mu = -5.0$, and $\Delta\mu/\mu = 0.05$.
Next, we address the case where the other affinity vanishes, which requires a precise relation between 
the temperature and chemical potential of the reservoirs $
\nabla\left(\frac{-\mu}{T}\right) = -\frac{\nabla\mu}{T} + \frac{\mu}{T^{2}}\nabla T = \frac{\mu}{N T}\left(-\frac{\Delta\mu}{\mu} + \frac{\Delta T}{T}\right) = 0$.
Our numerical calculations consider various mean temperature values $T = 0.5,\dots , 6.0$ and a mean chemical potential $\mu = -5$, 
such that $\Delta T/T = -0.05$ and $\Delta\mu/\mu = -0.05$, from where we extract the power and energy currents 
$j_a^{(\mu/T)}=\mathcal{L}^\prime_{ah}\cdot \nabla \left( \frac{1}{T}\right)$ and 
$j_h^{(\mu/T)}=\mathcal{L}^\prime_{hh}\cdot \nabla \left( \frac{1}{T}\right)$.

Finally, by using the definitions of the transport coefficients \cref{Eq:SupTransportCoeffs}, we calculate the WF ratio
\begin{equation}\label{Eq:SupWFnum}
\frac{\varkappa}{\sigma}  
= \frac{1}{T}\frac{\mathcal{L}^{\prime}_{hh}}{\mathcal{L}^{\prime}_{aa}}\left(1 - \frac{\mathcal{L}^{\prime}_{ah}\mathcal{L}^{\prime}_{ha}}{\mathcal{L}^{\prime}_{hh}\mathcal{L}^{\prime}_{aa}}\right)
=\frac{|\mu|}{T} \frac {j_{h}^{(\mu/T)}} {j_{a}^{(T)}} \left(1 - \frac{j_{a}^{(\mu/T)} \cdot j_{h}^{(T)}}{j_{a}^{(T)}\cdot j_{h}^{(\mu/T)}}\right),
\end{equation}
where we have used $\Delta\mu/|\mu| = \Delta T/T$. We have used this protocol and \cref{Eq:SupWFnum} to calculate all data points in Fig. (3) of the main text.



\end{document}